# A Formal Foundation for ODRL*

Riccardo Pucella
Northeastern University
Boston, MA 02115 USA
riccardo@ccs.neu.edu

Vicky Weissman
Cornell University
Ithaca, NY 14853 USA
vickyw@cs.cornell.edu

**Abstract**

ODRL is a popular XML-based language for stating the conditions under which resources can be accessed legitimately. The language is described in English and, as a result, agreements written in ODRL are open to interpretation. To address this problem, we propose a formal semantics for a representative fragment of the language. We use this semantics to determine precisely when a permission is implied by a set of ODRL statements and show that answering such questions is a decidable NP-hard problem. Finally, we define a tractable fragment of ODRL that is also fairly expressive.

## 1 Introduction

ODRL, the Open Digital Rights Language [Iannella 2002], is an XML-based language for stating the conditions under which resources can be accessed legitimately. For example, in ODRL, an author can write "Anyone who pays five dollars may download my latest eBook 'How to Get Rich in Five Dollar Increments' ". As another example, Pixar can say "The Disney Corp. has the exclusive right to distribute the movie 'Finding Nemo' ". Although there are many languages that can capture these types of statements, ODRL is particularly interesting because it has been endorsed by nearly twenty organizations including

- Nokia, a multi-industry conglomerate focused on mobile communications;

- the DAFNE project (District Architecture for Networked Editions), a research project funded by the Italian Ministry of Education, University and Research to develop a prototype of the national infrastructure for electronic publishing in Italy;

- the RoMEO Project (Rights MEtadata for Open archiving), created to investigate rights management of "self-archived" research in the United Kingdom academic community.

ODRL developers are currently working with a number of communities, including Creative Commons and Dublin Core, to address their needs. The complete list of supporters and on-going projects can be found at `www.odrl.net`; however, this small sample already illustrates the widespread impact that ODRL has on rights management. The success of these projects depends on ODRL.

---

*Most of this work was done while the first author was at Cornell University. A preliminary version of this paper was presented at the *Workshop on Issues in the Theory of Security (WITS'04)*, 2004.

Unfortunately, ODRL does not have formal semantics. The meaning of the statements is described in English and, as a result, agreements written in ODRL are open to interpretation. For example, suppose that Alice owns two printers, Printer One and Printer Two, and Bob is a potential user. To regulate Bob's access to the printers, Alice and Bob write an agreement in ODRL that says only this: Bob is permitted to use Printer One or Bob is permitted to use Printer Two. The agreement clearly allows Bob to use at least one of the printers, but it does not say which one. If Alice assumes the choice is hers, since the agreement does not say otherwise, and Bob believes the choice is his, since the agreement arguably implies this, then Alice and Bob disagree on the meaning of the agreement. Moreover, because this type of underspecification is possible in ODRL, they cannot use the ODRL specification to resolve the dispute.

To address this problem, we propose a formal semantics for ODRL and define when a permission (or prohibition) follows from a set of ODRL statements. To the best of our knowledge, we are the first to do this. When giving the language formal semantics, we had to resolve the ambiguities in the specification. Most of the aspects were clarified through discussions with Renato Iannella, editor of the ODRL specification and Chief Scientist at IPR systems at the time of its release. Unfortunately, he could not answer all of our questions because some of them revealed subtleties in the language that had not been considered previously. While discovering such subtleties is one of the rewards for trying to give a language formal semantics, these issues must be resolved before semantics can be given. So, when necessary, we have highlighted ambiguities and then taken our best guess.

We give formal semantics to ODRL by defining a translation from the key components in ODRL to formulas in a fragment of many-sorted first-order logic with equality. We use first-order logic because the formal methods community has proposed several policy languages that are fragments of first-order logic (see, for example, Cassandra [Becker and Sewell 2004], Lithium [Halpern and Weissman 2003], Delegation Logic [Li et al. 2003], the RT (Role-based Trust-management) framework [Li et al. 2002], Binder [DeTreville 2002], SD3 [Jim 2001], and FAF (Flexible Authorization Framework) [Jajodia et al. 2001]), and a translation exists for XrML [Halpern and Weissman 2004], another popular XML-based language. So the translation from ODRL to first-order logic facilitates comparisons between the languages and helps us apply previous results to ODRL. In addition, because first-order logic is highly expressive, we are hopeful that, if ODRL is extended, then the translation can be extended in a natural way.

The formal semantics can be used as a foundation for answering queries. For example, answering a query of the form "Does a particular permission (or prohibition) follow from a set of ODRL statements" corresponds to deciding whether the translation of the statements implies the permission (or prohibition). Answering this particular type of query is of obvious practical importance. Unfortunately, we show that the problem is NP-hard. The intractability result is due, at least in part, to a component that is not clearly defined in the specification and seems to require further consideration by the language developers. If we remove this troublesome construct, then we can answer our queries in polynomial time.

The rest of this paper is organized as follows. In the next section, we present a representative fragment of ODRL. In Section 3, we give a semantics to this fragment by translating expressions in the language to formulas in first-order logic. In Section 4, we define when a set of ODRL statements imply a permission (or prohibition); show that determining whether a particular implication holds is, in general, NP-hard; and find a tractable fragment of the language. We give a general critique of ODRL, along with suggested improvements, in Section 5. We conclude in Section 6.

## 2 The ODRL Language

In this section, we describe ODRL by giving an *abstract syntax* for a representative fragment of the language. Using this abstract syntax, rather than the XML-based syntax of ODRL, simplifies the presentation and discussion of our semantics. To illustrate the differences between the two notations, consider the statement "If Mary Smith pays five dollars, then she is allowed to print the eBook 'Treasure Island' twice and she is allowed to display it on her computer as many times as she likes". (A similar expression is discussed in [Guth et al. 2003].) We can write the statement in ODRL as

```
<agreement>
  <asset> <context> <uid> Treasure Island </uid> </context> </asset>
  <permission>
    <display>
      <constraint>
        <cpu> <context> <uid> Mary's computer </uid> </context> </cpu>
      </constraint>
    </display>
    <print>
      <constraint> <count> 2 </count> </constraint>
    </print>
    <requirement>
      <prepay>
        <payment> <amount currency="AUD"> 5.00</amount> </payment>
      </prepay>
    </requirement>
  </permission>
  <party> <context> <name> Mary Smith </name> </context> </party>
</agreement>
```

In our syntax, we write the statement as

> **agreement**
>    **for** *Mary Smith*
>    **about** *Treasure Island*
>    **with** prePay[5.00] $\longrightarrow$ and[cpu[*Mary's Computer*] $\Longrightarrow$ display,
>                                              count[2] $\Longrightarrow$ print].

Our syntax is given in Figures 1 and 2. We now discuss its main features and then present a summary of the key differences between our syntax and that of ODRL.

The central construct of ODRL is an *agreement*. An agreement says that a principal (i.e., an agent or a group) $prin_u$ is allowed to access an asset according to a set of policies (i.e., rules). Typically, $prin_u$ is called the agreement's user. For example, suppose that an agreement says "Alice is allowed to play 'Finding Nemo', if she first pays five dollars". Then, the user is Alice, the asset is 'Finding Nemo', and the policy is "The user may play the asset, if she pays five dollars"

| | |
|---|---|
| *agr* ::= | agreement |
|    **agreement** | |
|      **for** $prin_u$ | |
|      **about** $a$ | |
|      **with** $ps$ | |
| *prin* ::= | principal |
|    $s$ | subject |
|    $\{prin_1, \ldots, prin_m\}$ | group |
| $a \in \textit{Assets}$ | asset |
| $s \in \textit{Subjects}$ | subject |
| *ps* ::= | policy set |
|    $prq \longrightarrow p$ | primitive policy set |
|    $prq \longmapsto p$ | primitive exclusive policy set |
|    $\mathsf{and}[ps_1, \ldots, ps_m]$ | conjunction ($m \geq 1$) |
| *p* ::= | policy |
|    $prq \Longrightarrow_{id} act$ | primitive policy |
|    $\mathsf{and}[p_1, \ldots, p_m]$ | conjunction ($m \geq 1$) |
| *act* ::= | action |
|    $\mathsf{play}$ | play asset |
|    $\mathsf{print}$ | print asset |
|    $\mathsf{display}$ | display asset |
| $id \in \textit{PolIds}$ | policy identifier |

Figure 1: Abstract syntax for ODRL (agreements)

The set of principals and assets is application-dependent. For example, a digital library might have a principal for each patron and an asset for each publication. We assume that the application provides a set *Assets* of assets, as well as a set *Subjects* of subjects. The set of principals is defined inductively: every subject in *Subjects* is a principal and every group (i.e., set) of principals is a principal. Roughly speaking, if a policy applies to a principal *prin*, then the policy applies to every subject in *prin*.

Every agreement includes a *policy set*. A policy set consists of a prerequisite and a policy. Roughly speaking, if the prerequisite holds, then the policy holds; that is, the policy is taken into consideration when answering questions about what is and what is not permitted. In addition, a policy set can be tagged as being *exclusive*. An exclusive policy set indicates that only the agreement's user (the subjects comprising the principal) may perform the actions regulated by the policy set; every other subject is forbidden from doing the regulated actions. Policy sets are closed under conjunction. Roughly speaking, this allows a single agreement to include multiple policy sets.

A *policy* is a prerequisite, an action, and a unique identifier. If the prerequisite holds, then the policy says that the agreement's user may perform the action to the agreement's asset. (We use the identifiers to simplify the translation. They are optional in ODRL.) The set of policies is closed under conjunction. For simplicity, we often omit the identifier if it is not relevant to our examples and we restrict the set of actions to play, print, and display.

| | |
|---|---|
| $prq ::=$ | prerequisite |
|     true | |
|     $cons$ | constraint |
|     $req$ | requirement |
|     $cond$ | condition |
|     and$[prq_1, \ldots, prq_m]$ | conjunction $(m \geq 1)$ |
|     or$[prq_1, \ldots, prq_m]$ | disjunction $(m \geq 1)$ |
|     xor$[prq_1, \ldots, prq_m]$ | exclusive disjunction $(m \geq 1)$ |
| $cons ::=$ | constraint |
|     $prin$ | principal |
|     forEachMember$[prin; cons_1, \ldots, cons_m]$ | constraint distribution $(m \geq 1)$ |
|     count$[n]$ | number of executions $(n \in \mathbb{N})$ |
|     $prin\langle$count$[n]\rangle$ | number of executions by $prin$ $(n \in \mathbb{N})$ |
| $req ::=$ | requirement |
|     prePay$[r]$ | prepayment $(r \in \mathbb{R}_+)$ |
|     attribution$[s]$ | attribution to subject $s$ |
|     inSeq$[req_1, \ldots, req_m]$ | ordered constraints $(m \geq 1)$ |
|     anySeq$[req_1, \ldots, req_m]$ | unordered constraints $(m \geq 1)$ |
| $cond ::=$ | condition |
|     not$[ps]$ | suspending policy set |
|     not$[cons]$ | suspending constraint |

Figure 2: Abstract syntax for ODRL (prerequisites)

A prerequisite is either true, a constraint, a requirement, or a condition. The prerequisite true always holds. For simplicity, we abbreviate policy sets of the form true $\longrightarrow p$ as $p$, and we abbreviate policies of the form true $\Longrightarrow act$ as $act$. Constraints are facts that are outside the user's influence. For example, there is nothing that Alice can do to meet the constraint "The user is Bob". Requirements are facts that are typically within the user's power to meet. For example, Alice can meet the requirement "The user has paid five dollars" by making the payment. Although the distinction between constraints and requirements is not relevant when answering questions about what is and is not permitted, we remark that it is useful for other types of queries. In particular, it provides key information when determining what a principal can do to obtain a permission. Finally, conditions are constraints that must not hold. The statement "The user is not Bob" is an example of a condition.

The set of prerequisites is closed under conjunction, disjunction, and exclusive disjunction (i.e., under and, or, and xor). Conjunction allows a single policy or policy set to have multiple prerequisites. For example, we use conjunction to write the policy "If the user pays one dollar *and* acknowledges Alice as the creator of file $f$, then the user may copy $f$". Disjunction and exclusive disjunction are used to abbreviate policies and policy sets in a natural way. For example, consider the policy "If the user pays five dollars then the user may watch the movie *and* if the user is Alice, then the user may watch the movie". Using disjunction, we can abbreviate the policy as "If the user pays five dollars *or* the user is Alice, then the user may watch the movie".

Our fragment of ODRL includes two primitive forms of constraints *user constraints* and *count constraints*. A *user constraint* is a principal *prin*; a subject $s$ meets the constraint if $s \in prin$. A *count constraint* refers to a set $P$ of policies, and is parameterized by an integer $n$. The constraint holds if $n$ is greater than the number of times the user of the agreement has invoked the policies in $P$ to justify her actions. If the constraint appears in a policy $p$, then $P = \{p\}$. Otherwise, the constraint appears in some policy set $ps$ and $P$ is the set of policies mentioned in $ps$.

**Example 2.1.** Consider the following agreement:

$$\textbf{agreement for } \{Alice, Bob\} \textbf{ about } \textit{The Report} \textbf{ with } \mathsf{and}[p_1, p_2],$$

where $p_1$ is $\mathsf{count}[5] \Longrightarrow_{id_1} \mathsf{print}$ and $p_2$ is $\mathsf{and}[Alice, \mathsf{count}[2]] \Longrightarrow_{id_2} \mathsf{print}$. (Recall that $\mathsf{and}[p_1, p_2]$ is an abbreviation for the policy set $\mathsf{true} \longrightarrow \mathsf{and}[p_1, p_2]$.) The agreement says that asset *The Report* may be printed a total of five times by either Alice or Bob, and twice more by Alice. That is, if Alice and Bob have used policy $p_1$ to justify their printing of *The Report* $a_1$ and $b_1$ times, respectively, then either may do so again if $a_1 + b_1 < 5$. Similarly, if Alice and Bob have used the policy $p_2$ to justify printing $a_2$ and $b_2$ times, respectively, then Alice may do so again if $a_2 + b_2 < 5$. Note that, since Bob does not satisfy the constraint of being Alice, $b_2$ is 0, so the second policy amounts to giving Alice the permission to print *The Report* twice (in addition to any printings made by invoking other policies). ∎

A count constraint that appears in a policy set is interpreted in a similar way.

**Example 2.2.** Consider the following agreement:

$$\textbf{agreement for } \{Alice, Bob\} \textbf{ about } \textit{The Report} \textbf{ with } \mathsf{count}[5] \longrightarrow \mathsf{and}[p_1, p_2],$$

where $p_1$ is $\mathsf{print}$ and $p_2$ is $\mathsf{display}$. The agreement says that Alice and Bob may invoke policies $p_1$ and $p_2$ a total of five times to justify the printing or displaying of asset *The Report*. That is, if Alice and Bob have used policy $p_1$ to justify the $\mathsf{print}$ action $a_p$ and $b_p$ times respectively, and have used policy $p_2$ to justify the $\mathsf{display}$ action $a_d$ and $b_d$ times respectively, then either of them may print or display again if $a_p + b_p + a_d + b_d < 5$. ∎

The constraint $\mathsf{forEachMember}$ takes a principal *prin* (usually a group) and a list $L$ of constraints; it holds if each principal in *prin* satisfies each constraint in $L$.

ODRL supports nested constraints, where a constraint is used to modify another constraint. To illustrate how our approach can accommodate nested constraints, we support the constraint $prin\langle\mathsf{count}[n]\rangle$, which is interpreted like a $\mathsf{count}[n]$ constraint, except that it applies to the principal *prin* rather than to the user of the agreement. Thus, the constraint holds if $n$ is greater than the number of times *prin* has used the policies to justify her actions.

**Example 2.3.** Consider the following agreement:

$$\textbf{agreement for } \{Alice, Bob\} \textbf{ about } \textit{The Report} \textbf{ with } ps,$$

where $ps$ is $\mathsf{true} \longrightarrow p$ and $p$ is $Alice\langle\mathsf{count}[1]\rangle \Longrightarrow \mathsf{print}$. The agreement says that if Alice has not invoked policy $p$ to print asset *The Report*, then she may do so; until she does, Bob may use $p$ to print *The Report* any number of times. ∎

**Example 2.4.** Consider the following agreement:

$$\textbf{agreement for } \{Alice, Bob, Charlie\} \textbf{ about } \textit{The Report} \textbf{ with } ps,$$

where $ps$ is $\mathsf{and}[\{Alice, Bob\}, \{Alice, Bob\}\langle\mathsf{count}[5]\rangle] \longrightarrow \mathsf{and}[p_1, p_2]$, $p_1$ is $\mathsf{print}$, and $p_2$ is $\mathsf{display}$. The agreement says that Alice and Bob may invoke policies $p_1$ and $p_2$ a total of five times to justify printing and displaying asset *The Report*. Since Charlie does not satisfy the prerequisite $\{Alice, Bob\}$, he cannot invoke $p_1$ or $p_2$. ∎

There are two primitive requirements, $\mathsf{prePay}$ and $\mathsf{attribution}$. The $\mathsf{prePay}$ requirement takes an amount of money as a parameter; it holds if the user pays the specified amount. The $\mathsf{attribution}$ requirement takes a subject $s$ as a parameter; it holds if $s$ is properly acknowledged (e.g., as the writer, producer, etc.). The set of requirements is closed under the $\mathsf{inSeq}$ construct, which says the requirements must be met in a particular order (e.g., acknowledge, then pay), and under the $\mathsf{anySeq}$ construct, which says the requirements can be met in any order.

Finally, there are two types of conditions, negated constraints and negated policy sets. The condition $\mathsf{not}[cons]$ holds if and only if the constraint *cons* does not hold. For example, $\mathsf{not}[Alice]$ holds if and only if the user is not Alice. Similarly, the condition $\mathsf{not}[ps]$ holds if and only if the policy set $ps$ does not hold. But what does it mean that a policy set (or, in particular, a policy) does not hold? Consider the policy "If Alice pays five dollars, then she is permitted to play 'Finding Nemo'". There are at least two reasonable interpretations of when the policy does not hold. Under the first interpretation, the policy does not hold if Alice cannot get the permission by paying five dollars. In other words, we could interpret $\mathsf{not}[ps]$ to mean that a certain set of agreements does not imply $ps$. A problem with this interpretation is that we do not know which agreements should be used to evaluate the condition. Under the second interpretation, which we favor, the policy does not hold if Alice has paid five dollars and is not permitted to play the movie. In other words, the condition amounts to the logical negation of the policy. We choose this interpretation because it is simple, fairly intuitive, and, as we shall see, leads to semantics that matches the semantics for negated constraints. (This is encouraging because, in the ODRL specification, the discussion of negated policy sets is essentially identical to the discussion of negated constraints.)[1]

**Example 2.5.** Consider the following agreement:

---

[1]It is worth noting that we could modify our interpretation without contradicting the specification. Continuing with our example, one variation is to have the condition hold if Alice paid five dollars and is not explicitly permitted to play the movie. Another variation is to have the condition hold if Alice paid five dollars and is explicitly forbidden to play the movie. We could modify our semantics to accommodate the variations in a fairly straightforward way. (This can be accomplished with a validity operator; see [Halpern and Weissman 2004] for some details.)

**agreement**
　　**for** $\{Alice, Bob\}$
　　**about** *ebook*
　　**with** count[10] $\longrightarrow$ and[forEachMember[$\{Alice, Bob\}$; count[5]] $\Longrightarrow_{id_1}$ display,
　　　　　　　　　　　forEachMember[$\{Alice, Bob\}$; count[1]] $\Longrightarrow_{id_2}$ print].

The agreement says that Alice and Bob may each display the asset *ebook* up to five times, and they may each print it once. However, the total number of actions, either displays or prints, done by Alice and Bob may be at most ten. ∎

**Example 2.6.** Consider the following agreement:

**agreement**
　　**for** $\{Alice, Bob\}$
　　**about** *latestJingle*
　　**with** inSeq[prePay[5.00], attribution[$Charlie$]] $\longmapsto$ ($Alice\langle$count[10]$\rangle \Longrightarrow_{id}$ play).

The agreement says that after paying five dollars and then acknowledging Charlie, Alice is permitted to play the asset *latestJingle* up to ten times. Moreover, any subject that is neither Alice nor Bob is forbidden from playing *latestJingle*. (Bob's right is unregulated.) ∎

　　As mentioned at the beginning of this section, the syntax presented here differs from the one described in the ODRL specification. The key differences are discussed below.

　　**Authorship.** An ODRL agreement includes a principal called the owner. Roughly speaking, the owner is the principal who is granting the permissions that are mentioned in the agreement. While this information can be useful in practice (e.g., for auditing), our syntax does not mention the owner of an agreement because the identity of the owner does not affect the legitimacy of an ODRL agreement—an agreement holds regardless of who created it.

　　**Offers.** In addition to agreements, ODRL includes offers, which are essentially agreements without users. Intuitively, an offer is a contract (governing the use of an asset) that does not apply until it is accepted by a user; once accepted, it becomes an agreement. We can interpret offers much as we do agreements.

　　**Permissions versus Policies.** The ODRL specification uses the term *permission* to refer to actions, policies, and policy sets, as defined here. We introduce the distinction to clarify the exposition and to emphasize the two-tier structure of ODRL. Notice that it is the two layers in the framework that allow a prerequisite to apply to multiple policies.

　　**Contexts.** ODRL uses contexts to assign additional information to agreements, prerequisites, and other entities. A context might include a unique identifier, a human-readable name, an expiration date, and so on. We represent the context elements that are included in our fragment directly in the syntax. Adding full contexts to our syntax is straightforward, but it does not add any insight. Moreover, we believe it obfuscates the main issues.

　　**Prerequisites.** Payments and other requirements in ODRL take a number of arguments. For instance, payments can take an amount and a percentage to be collected for taxes. We restrict every prerequisite to at most one argument for simplicity; it is easy to extend our approach to include multiple arguments. As we have already mentioned, ODRL supports nested constraints. These can be handled in a manner similar to that used for $prin\langle$count[n]$\rangle$.

**Sequences and Containers.** In ODRL, sequences (inSeq, anySeq) and containers (and, or, xor) apply to a number of entities. For simplicity, we associate the three containers with prerequisites, and associate sequences with requirements. The general case is a straightforward extension. In particular, the extension of containers to policies in the obvious way helps resolve the ambiguity discussed in the introduction; the policy "Bob may use Printer One or Bob may use Printer Two" gives Bob the right to use either printer as he chooses. According to discussions with Renato Iannella, this is the interpretation intended by the language developers.

**Right Holders.** In ODRL, right holders have a royalty annotation, indicating the amount of royalty that they receive. This does not reflect an obligation on the part of the agreement's user, since payment obligations are captured by requirements. Instead, the annotations record how the payments are distributed. Since we are primarily interested in capturing permissions, we do not consider royalty annotations, and as a result, do not distinguish right holders from other principals.

**Revocation.** Finally, the ODRL specification mentions revocation, however it is not clearly defined. A revocation invalidates a previously established agreement. Unfortunately, answers to key questions, such as who can revoke an agreement, under what conditions, and subject to what penalties, are not discussed in the ODRL specification. As it stands, a revocation simply indicates that an agreement has been nullified, and thus may be ignored.

# 3 A Semantics in First-Order Logic

In this section, we formalize the intuitive description of ODRL given in Section 2. Specifically, we present a translation from agreements to formulas in many-sorted first-order logic with equality. For the rest of this discussion, we assume knowledge of many-sorted first-order logic at the level of Enderton [1972]. More specifically, we assume familiarity with the syntax of first-order logic, including constants, variables, predicate symbols, function symbols, and quantification, with the semantics of first-order logic, including relational models and valuations, and with the notion of satisfiability and validity of first-order formulas.

We assume sorts *Actions*, *Subjects*, *Assets*, *PolIds*, and *SetPolIds* (for sets of policy identifiers), and deliberately identify a sort with the set of values of that sort. We further assume sorts *Reals* and *Times*; *Real* to represent real numbers, and *Times* to represent time. We interpret real numbers in the standard way. For simplicity, we take sort *Times* to be the nonnegative real numbers extended with the special constant $\infty$ representing infinity. Again, we interpret such extended nonnegative real numbers in the standard way; in particular, $t < \infty$ for every nonnegative real number $t$ different from $\infty$.

The vocabulary includes:

- A predicate **Permitted** on *Subjects* $\times$ *Actions* $\times$ *Assets*. The literal **Permitted**$(s, act, a)$ means $s$ is permitted to perform action $act$ on asset $a$.

- A predicate **Paid** on *Reals* $\times$ *SetPolIds* $\times$ *Time*. The literal **Paid**$(r, I, t)$ means an amount $r$ was paid towards the policies corresponding to the set $I$ of policy identifiers at time $t$.

- A predicate **Attributed** on *Subjects* × *Time*. The literal **Attributed**$(s, t)$ means $s$ was acknowledged at time $t$.

- Constants of sort *PolIds*, *SetPolIds*, *Subjects*, and *Assets*; we also assume constants *play*, *display*, and *print* of sort *Actions*.

- A function *count* : *Subjects* × *PolIds* → *Reals*. Intuitively, $count(s, id)$ is the number of times subject $s$ used the policy with identifier $id$ to justify an action.

- Standard functions for addition $(+)$ and comparison $(<, \leq)$ of real numbers and extended real numbers.

Before presenting the translation, we define some useful auxiliary functions. The function *subjects* returns the set all subjects appearing in a principal:

$$subjects(s) \triangleq \{s\}$$

$$subjects(\{prin_1, \ldots, prin_k\}) \triangleq \cup_{i=1}^{k} subjects(prin_i).$$

The function *principals* returns the set of principals that are members of a given principal; if the principal is a subject, the function returns the singleton set consisting of that subject:

$$principals(s) \triangleq \{s\}$$

$$principals(\{prin_1, \ldots, prin_k\}) \triangleq \{prin_1, \ldots, prin_k\}.$$

The function *ids* takes a policy $p$, and returns the set of policy identifiers that are mentioned in $p$:

$$ids(pr_1 \ \ldots \ pr_m \Longrightarrow_{id} act) \triangleq \{id\}$$

$$ids(\mathsf{and}[p_1, \ldots, p_m]) \triangleq \bigcup_{i=1}^{m} ids(p_i).$$

The translation proceeds by induction on the structure of the agreement. The translation is given in Figures 3 and 4; we discuss its key features below.

An agreement is translated into a conjunction of formulas of the form:

$$\forall x (prerequisites(x) \Rightarrow P(x)),$$

where $P(x)$ is itself a conjunction of formulas of the form

$$prerequisites(x) \Rightarrow (\neg)\mathbf{Permitted}(x, act, a)$$

and $x$ is a variable of sort *Subjects* that is free in $P(x)$. (The notation $(\neg)\mathbf{Permitted}(\cdot)$ indicates that the formula $\mathbf{Permitted}(\cdot)$ might be negated.)

The translation of a policy set $ps$ is a formula $[\![ps]\!]^{prin_u, a}$, where $prin_u$ is the agreement's user and $a$ is the asset. A (nonexclusive) primitive policy set $prq \longrightarrow p$ translates to an implication: if the user is in $prin_u$ and the prerequisite holds, then the policy holds. An exclusive primitive policy set is translated as a nonexclusive primitive policy set in conjunction with a clause that captures the prohibition (i.e., every subject that is not $[\![\textbf{agreement for } prin_u \textbf{ about } a \textbf{ with } ps]\!] \triangleq [\![ps]\!]^{prin_u,a}$

$[\![prq \longrightarrow p]\!]^{prin_u,a} \triangleq \forall x(([\![prin_u]\!]_x \wedge [\![prq]\!]_x^{ids(p),prin_u,a}) \Rightarrow [\![p]\!]_x^{+,prin_u,a})$

$[\![prq \longmapsto p]\!]^{prin_u,a} \triangleq \forall x(([\![prin_u]\!]_x \wedge [\![prq]\!]_x^{ids(p),prin_u,a}) \Rightarrow [\![p]\!]_x^{+,prin_u,a})$
$\qquad\qquad\qquad \wedge \forall x(\neg[\![prin_u]\!]_x \Rightarrow [\![p]\!]_x^{-,a})$

$[\![\textsf{and}[ps_1,\ldots,ps_m]]\!]^{prin_u,a} \triangleq \bigwedge_{i=1}^m [\![ps_i]\!]^{prin_u,a}$

$[\![s]\!]_x \triangleq x = s$
$[\![\{prin_1,\ldots,prin_k\}]\!]_x \triangleq ([\![prin_1]\!]_x \vee \ldots \vee [\![prin_k]\!]_x)$

$[\![prq \Longrightarrow_{id} act]\!]_x^{+,prin_u,a} \triangleq ([\![prq]\!]_x^{\{id\},prin_u,a}) \Rightarrow \textbf{Permitted}(x,[\![act]\!],a)$
$[\![\textsf{and}[p_1,\ldots,p_m]]\!]_x^{+,prin_u,a} \triangleq \bigwedge_{i=1}^m [\![p_i]\!]_x^{+,prin_u,a}$

$[\![prq_1 \; \ldots \; prq_m \Longrightarrow_{id} act]\!]_x^{-,a} \triangleq \neg\textbf{Permitted}(x,[\![act]\!],a)$
$[\![\textsf{and}[p_1,\ldots,p_m]]\!]_x^{-,a} \triangleq \bigwedge_{i=1}^m [\![p_i]\!]_x^{-,a}$

$[\![\textsf{play}]\!] \triangleq play$
$[\![\textsf{display}]\!] \triangleq display$
$[\![\textsf{print}]\!] \triangleq print$

Figure 3: Translation of ODRL agreements

mentioned in the agreement's user is forbidden from performing the actions). Conjunctions of policy sets translate to conjunctions of the corresponding formulas. (In the translation, we follow the convention that $\bigwedge_{i=1}^m f_i$ is **true** when $m = 0$.) Note that the translation of a policy set is defined in terms of a check that the user is in $prin_u$, the translation of a policy, and the translation of a prerequisite. We now consider each of these in turn. The formula $[\![prin]\!]_x$ is true if and only if the subject denoted by the variable $x$ is in the principal $prin$.

There are two translations for policies: a positive translation, where the permissions described by a policy are granted, and a negative translation, where they are forbidden. The positive translation of a policy $p$ is a formula $[\![p]\!]_x^{+,prin_u,a}$, where $prin_u$ is the user of the agreement, $a$ is the asset, and $x$ is the variable that ranges over the subjects. A policy of the form $prq \Longrightarrow act$ translates to an implication: if the prerequisite holds, then the subject represented by $x$ is permitted to perform the action $act$ on the asset $a$. The negative translation of a policy $p$ is a formula $[\![p]\!]_x^{-,a}$, where $a$ is the asset, and $x$ is the variable that ranges over the subjects. If $p$ is $prq \Longrightarrow act$, then the translation says that $x$ is forbidden to do $act$ to $a$, regardless of whether $prq$ holds. The positive and negative translations of policies are defined in terms of the translation of actions, which is simply the constant corresponding to the action. As with policy sets, conjunctions of policies translate to conjunctions of the corresponding formulas.

The translation of a prerequisite $prq$ is a formula $[\![prq]\!]_x^{I,prin,a}$, where $I$ is a set of policy identifiers, $prin$ is a principal, $a$ is an asset, and $x$ is a variable of sort *Subjects*. Intuitively,

$$\llbracket \mathsf{true} \rrbracket_x^{I,prin_u,a} \triangleq \mathbf{true}$$

$$\llbracket prin \rrbracket_x^{I,prin_u,a} \triangleq \llbracket prin \rrbracket_x$$

$$\llbracket \mathsf{forEachMember}[prin; cons_1, \ldots, cons_m] \rrbracket_x^{I,prin_u,a} \triangleq \bigwedge_{(prin',i) \in P_m} \llbracket cons_i \rrbracket_x^{I,prin',a}$$

   where $P_m = principals(prin) \times \{1, \ldots, m\}$

$$\llbracket \mathsf{count}[n] \rrbracket_x^{I,prin_u,a} \triangleq \left( \sum_{(id,s) \in I \times (subjects(prin_u))} count(s, id) \right) < n$$

$$\llbracket prin \langle \mathsf{count}[n] \rangle \rrbracket_x^{I,prin_u,a} \triangleq \left( \sum_{(id,s) \in I \times (subjects(prin))} count(s, id) \right) < n$$

$$\llbracket req \rrbracket_x^{I,prin_u,a} \triangleq \llbracket req \rrbracket_{0,\infty}^I$$

   where $\llbracket \mathsf{prePay}[r] \rrbracket_{t,t'}^I \triangleq \exists t''(t \le t'' < t' \wedge \mathbf{Paid}(r, I, t''))$

   $\llbracket \mathsf{attribution}[s] \rrbracket_{t,t'}^I \triangleq \exists t''(t \le t'' < t' \wedge \mathbf{Attributed}(s, t''))$

   $\llbracket \mathsf{inSeq}[req_1, \ldots, req_k] \rrbracket_{t,t'}^I \triangleq$

   $\begin{cases} \llbracket req_1 \rrbracket_{t,t'}^I & \text{if } k = 1 \\ \exists t_2 \ldots \exists t_k (t < t_2 < \cdots < t_k < t' \wedge \llbracket req_1 \rrbracket_{t,t_2}^I \wedge \cdots \wedge \llbracket req_k \rrbracket_{t_k,t'}^I) & \text{if } k \ge 2 \end{cases}$

   $\llbracket \mathsf{anySeq}[req_1, \ldots, req_k] \rrbracket_{t,t'}^I \triangleq \bigwedge_{i=1}^k \llbracket req_i \rrbracket_{t,t'}^I$

$$\llbracket \mathsf{not}[ps] \rrbracket_x^{I,prin_u} \triangleq \neg(\llbracket ps \rrbracket^{prin_u,a})$$

$$\llbracket \mathsf{not}[cons] \rrbracket_x^{I,prin_u} \triangleq \neg \llbracket cons \rrbracket_x^{I,prin_u}$$

$$\llbracket \mathsf{and}[prq_1, \ldots, prq_m] \rrbracket_x^{I,prin_u} \triangleq \bigwedge_{i=1}^m \llbracket prq_i \rrbracket_x^{I,prin_u}$$

$$\llbracket \mathsf{or}[prq_1, \ldots, prq_m] \rrbracket_x^{I,prin_u} \triangleq \bigvee_{i=1}^m \llbracket prq_i \rrbracket_x^{I,prin_u}$$

$$\llbracket \mathsf{xor}[prq_1, \ldots, prq_m] \rrbracket_x^{I,prin_u} \triangleq \bigvee_{i=1}^m (\llbracket prq_i \rrbracket_x^{I,prin_u} \wedge (\bigwedge_{j=1,j\neq i}^m \neg \llbracket prq_j \rrbracket_x^{I,prin_u}))$$

Figure 4: Translation of ODRL prerequisites

$I$ includes (the identifier of) the policies that are implied by the prerequisites and $prin$ is the principal to which the prerequisites apply (the agreement's user, unless overridden within a forEachMember constraint). A Boolean combination of prerequisites translates to the Boolean combination of the formulas obtained by translating each prerequisite in turn. A user constraint $prin$ translates to a formula that is true if the current subject $x$ is a member of $prin$. The translation of the other constraints is more complicated. A forEachMember constraint translates to a formula that is true if, intuitively, each constraint in forEachMember is met by each subject mentioned in the constraint (i.e., each member). A constraint count[$n$] translates to a formula that is true if the subjects mentioned in $prin_u$ have invoked the policies identified in $I$ a total of $i$ times where $i$ is less than $n$. Similarly, a $prin\langle$count[$n$]$\rangle$ constraint translates to a formula that is true if the total number of times that a subject in $prin$ has invoked a policy whose identifier is in $I$ is less than $n$.

Requirements have a significantly different translation than other prerequisites because of their dependence on time (e.g., inSeq[prePay[$r$], attribution[$s$]] holds if $r$ is paid before $s$ is acknowledged). To handle time correctly, we translate $\llbracket req \rrbracket_x^{I,prin,a}$ to $\llbracket req \rrbracket_{0,\infty}^I$, where $\llbracket req \rrbracket_{t,t'}^I$ is an auxiliary translation that returns a formula that is $\mathbf{true}$ if the events specified by requirement $req$ occur within the interval of time between $t$ and $t'$. If $req$ is a primitive requirement (i.e., a payment or attribution), then we translate $[\![req]\!]_{t,t'}^{I}$ to a formula that is true if the relevant payment or attribution occurred at some time between $t$ and $t'$. An inSeq requirement is satisfied if there exists appropriate successive times between $t$ and $t'$ at which each subrequirement is satisfied. Similarly, an anySeq requirement is satisfied if the subrequirements are satisfied in an arbitrary order (possibly simultaneously) between times $t$ and $t'$.

Conditions are translated by negating the translation of either the policy set or the constraint specified as the argument. Recall that, in ODRL, we can capture statements such as "If Alice is not permitted to print the report, then she is permitted to display it". We can also write "If Alice is permitted to print the report, then she is permitted to display it", since xor[true, not[$ps$]] is equivalent to $ps$. It follows from our semantics that the first statement alone gives Alice the display permission if she is explicitly forbidden to print the report; the two statements together imply that Alice may display the report, regardless of which print permissions are granted or denied.

Another subtlety arises in the interpretation of sequence requirements, particularly *nested* sequence requirements. To illustrate the issue, consider the nested requirement anySeq[inSeq[$req_1$, $req_2$], $req_3$]. What are the allowed sequences of requirements $req_1$, $req_2$, and $req_3$? One possibility, the one we adopt, is that inSeq[$req_1$, $req_2$] is met if $req_1$ happens before $req_2$. Thus, the following sequences are allowed: $req_1$ $req_2$ $req_3$, $req_1$ $req_3$ $req_2$, and $req_3$ $req_1$ $req_2$. Alternatively, one could say that inSeq[$req_1$, $req_2$] is met if $req_1$ and $req_2$ happen consecutively. Under this interpretation, only the following sequences are allowed: $req_1$ $req_2$ $req_3$ and $req_3$ $req_1$ $req_2$. We can capture this last interpretation by taking:

$$[\![\mathsf{anySeq}[req_1,\ldots,req_k]]\!]_{t,t'}^{I} \triangleq$$
$$\begin{cases} [\![req_1]\!]_{t,t'}^{I} & \text{if } k = 1 \\ \exists t_2 \ldots \exists t_k (t < t_2 < \cdots < t_k < t' \land \bigvee_{\pi \in S_k} ([\![req_{\pi(1)}]\!]_{t,t_2}^{I} \land \cdots \land [\![req_{\pi(k)}]\!]_{t_k,t'}^{I}) & \text{if } k \geq 2, \end{cases}$$

where $S_k$ is the set of all permutations of sets of $k$ elements.

Our translation is admittedly complex, however it is not clear that a more simple translation is possible due to the distributed nature of agreements (e.g., a count constraint can implicitly refer to policy identifiers that occur throughout the enclosing policy set). To conclude this section, we translate Examples 2.5 and 2.6 from Section 2.

**Example 3.1.** The agreement in Example 2.5

> **agreement**
>     **for** $\{Alice, Bob\}$
>     **about** *ebook*
>     **with** count[10] $\longrightarrow$ and[forEachMember[$\{Alice, Bob\}$; count[5]] $\Longrightarrow_{id_1}$ display,
>                              forEachMember[$\{Alice, Bob\}$; count[1]] $\Longrightarrow_{id_2}$ print]

translates to the formula

$\forall x((x = Alice \lor x = Bob) \Rightarrow$
$\quad count(Alice, id_1) + count(Alice, id_2) + count(Bob, id_1) + count(Bob, id_2) \leq 10 \Rightarrow$
$\quad\quad ((count(Alice, id_1) < 5 \land count(Bob, id_1) < 5) \Rightarrow$
$\quad\quad\quad \textbf{Permitted}(x, display, ebook)) \land$
$\quad\quad ((count(Alice, id_2) < 1 \land count(Bob, id_2) < 1) \Rightarrow$
$\quad\quad\quad \textbf{Permitted}(x, print, ebook))).$

∎

**Example 3.2.** The agreement in Example 2.6

> **agreement**
> **for** $\{Alice, Bob\}$
> **about** *latestJingle*
> **with** inSeq[prePay[5.00], attribution[*Charlie*]] $\longmapsto$ (*Alice*⟨count[10]⟩ $\Longrightarrow_{id}$ play)

translates to the formula

$\forall x((x = Alice \lor x = Bob) \Rightarrow$
$\quad \exists t_1 \exists t_2 (t_1 < t_2 \land \textbf{Paid}(5.00, t_1) \land \textbf{Attributed}(Charlie, t_2)) \Rightarrow$
$\quad\quad (x = Alice \land count(Alice, id) < 10 \Rightarrow \textbf{Permitted}(x, play, latestJingle)) \land$
$\quad\quad (\neg(x = Alice \lor x = Bob) \Rightarrow \neg\textbf{Permitted}(x, play, latestJingle))).$

∎

These examples illustrate that, despite the complexity of the translation, the structure of formulas obtained from the translation follows closely that of the agreements.

# 4 Queries

Our formal semantics provides a foundation for reasoning about agreements in a rigorous way. Because of their obvious usefulness, we focus on queries of the form "may subject $s$ do action $act$ to asset $a$". In this section, we formally define such queries; then we examine the complexity of answering them.

## 4.1 Formal Definition

Whether a permission (or prohibition) holds depends on the agreements that have been created, as well as certain facts about the application. For our fragment of ODRL, the relevant facts are which payments have been made, which acknowledgments have been given, and the number of times each policy has been used to justify an action. We encode this information in an *environment*, which is a conjunction of positive ground literals, each of the form $\textbf{Attributed}(s, t)$ or $\textbf{Paid}(s, I, t)$, and equalities of the form $count(s, id) = n$. Based on the type of information stored in the environment (both for our fragment and for all of ODRL), it seems reasonable to make a form of *closed-world assumption*: we assume all environment facts are known. That is, if a positive $\textbf{Permitted}$-free ground literal is not a conjunct of the environment then we assume it does not hold, with two exceptions. First, if there is a subject $s$ and policy identifier $id$ such that no conjunct of $E$ has the form $count(s, id) = n$, then we assume $count(s, id) = 0$. Second, if the environment together with the standard interpretation of $=$, $<$, and $\leq$ imply that a positive literal holds, then we assume that it does. For example, if $s$ and $s'$ are subjects; $id$ and $id'$ are policy identifiers; and no conjunct of $E$ has the form $count(s, id) = n$ or $count(s', id') = n$, then we assume $count(s, id) = 0$, $count(s', id') = 0$, and $count(s, id) = count(s', id')$.

Suppose that we are interested in determining whether a set $A$ of agreements imply that a subject $s$ may do action $act$ to asset $a$ in environment $E$. We represent such a query as a tuple $(A, s, act, a, E)$. Answering the query corresponds to establishing the validity of a formula with respect to a particular class of models. Recall that a *Herbrand* model is a model whose domain consists of the closed terms in the language. We are interested only in Herbrand models that agree with the environment and that interpret the symbols $=$, $<$, and $\leq$ in the standard way; that is, they satisfy the axioms of *real closed fields* [Tarski 1951] over the sorts *Reals* and *Times*—in the latter case, the axioms extended with the obvious axioms to deal with $\infty$. These axioms include, for instance, the reflexivity of equality, $\forall x.(x = x)$, and the monotonicity of addition, $\forall x, y, z.(x \leq y \Rightarrow x + z \leq y + z)$. Moreover, we want the models to enforce the closed-world assumption on environments. Given an environment $E$, let $\mathcal{F}(E)$ be the set of formulas made up of $E$ itself, the real closed fields axioms (extended to deal with $\infty$), and formulas $\mathsf{count}(s, id) = 0$ for every subject $s$ and policy identifier $id$ such that $\mathsf{count}(s, id)$ is not a conjunct of $E$. Intuitively, these are the formulas directly "implied" by the environment. Given a query $q = (A, s, act, a, E)$, define a model $M$ to be *E-relevant* if:

(1) the domain of $M$ consists of the closed terms in the language;

(2) $M$ satisfies every formula in $\mathcal{F}(E)$;

(3) for every positive **Permitted**-free ground literal $\ell$ that holds in $M$, the model $M'$ that is identical to $M$ except that it does not satisfy $\ell$ does not satisfy every formula in $\mathcal{F}(E)$.

Because an environment consists only of positive facts, an environment $E$ is inconsistent if and only if $E$ has two conjuncts $count(s, id) = n_1$ and $count(s, id) = n_2$ with $n_1 \neq n_2$. Thus, an environment $E$ is consistent if and only if there exists an $E$-relevant model. When evaluating a query $q = (A, s, act, a, E)$, we consider only those models that are *E-relevant*. A formula is *E-valid* if it holds in every $E$-relevant model.

We now have the necessary foundation to give an answer to a query $q = (A, s, act, a, E)$. Define the formulas:

$$f_q^+ \triangleq \bigwedge_{agr \in A} [\![agr]\!] \Rightarrow \mathbf{Permitted}(s, act, a)$$

$$f_q^- \triangleq \bigwedge_{agr \in A} [\![agr]\!] \Rightarrow \neg\mathbf{Permitted}(s, act, a).$$

The answer to the query depends on the $E$-validity of $f_q^+$ and $f_q^-$.

- If both $f_q^+$ and $f_q^-$ are $E$-valid, then either the environment is inconsistent, in which case all formulas are $E$-valid, or the agreements are inconsistent in the environment. Either way, an appropriate answer to the query seems to be "Query inconsistent".

- If $f_q^+$ is $E$-valid and $f_q^-$ is not, the answer is "Permission granted" because, roughly speaking, the permission necessarily follows from the agreements in the given environment.

- Similarly, if $f_q^-$ is $E$-valid and $f_q^+$ is not, then the answer is "Permission denied".

- Finally, if neither $f_q^+$ nor $f_q^-$ is valid, then the agreements in the given environment do not imply that the permission is granted, nor do they imply that the permission is denied. So the answer is "Permission unregulated".

## 4.2 Complexity

We now consider the computational complexity of answering queries. It turns out that we can create an algorithm that takes a query and returns the correct answer; however, it seems unlikely that any algorithm will run efficiently on all input. The relevant result is the following.

**Theorem 4.1.** *The problem of deciding, for a query $q = (A, s, act, a, E)$, whether $f_q^+$ is $E$-valid is decidable and NP-hard. Similarly, the problem of deciding, for a query $q = (A, s, act, a, E)$, whether $f_q^-$ is $E$-valid is decidable.*

*Proof.* See Appendix A. ∎

Since answering a query $q$ amounts to determining the $E$-validity of $f_q^+$ and $f_q^-$, the first of which cannot be done efficiently, answering a query cannot be done efficiently.

The proof of Theorem 4.1 in Appendix A suggests that the intractability result holds, at least in part, because ODRL includes conditions of the form $\mathsf{not}[ps]$, where $ps$ is a policy set. It might be possible to modify our translation, and thus the meaning, of $\mathsf{not}[ps]$ in such a way that the revised semantics matches the specification and answering queries in the revised language is a tractable (i.e., solvable in polynomial time) problem. This is because, as discussed in Section 2, the description of $\mathsf{not}[ps]$ in the ODRL specification is open to interpretation. However, tweaking the semantics to get a desired complexity result seems somewhat dubious. In addition, it is not clear that finding the largest tractable fragment of ODRL, as we have interpreted the language, is interesting because, having discovered that a component of the language is not clearly specified and a natural interpretation leads to intractability, it seems likely that the meaning of that component will be revised. Since we cannot know beforehand what the revision will be, we restrict our attention to the fragment of ODRL that does not include conditions of the form $\mathsf{not}[ps]$.

Let $\mathcal{Q}_1$ be the set of queries $(A, s, act, a, E)$ such that no agreement in $A$ mentions a prerequisite of the form $\mathsf{not}[ps]$. We now show that we can answer a query $q = (A, s, act, a, E)$ in $\mathcal{Q}_1$ efficiently. As a first step, we consider the special case in which the set of agreements is a singleton. For any expression $e$ (either in our ODRL syntax or in first-order logic), let $|e|$ be the length of $e$ when viewed as a string of symbols. For a set $A$ of agreements, let $|A|$ be $\Sigma_{agr \in A} |agr|$.

**Lemma 4.2.** *There are algorithms that, given a query $q = (\{agr\}, s, act, a, E)$ in $\mathcal{Q}_1$:*

(a) *determine whether $f_q^+$ is $E$-valid in time $O(|E| \, |agr|^6)$, and*

*(b) determine whether $f_q^-$ is E-valid in time $O(|E| + |agr|)$.*

*Proof.* See Appendix A. ∎

It follows from Lemma 4.2 that $\mathcal{Q}_1$ is tractable, provided that a permission (or prohibition) follows from a set of agreements if and only if it follows from a single agreement in the set. Unfortunately, this is not necessarily true.

**Example 4.3.** Let $A = \{agr, agr'\}$, where $agr$ is

**agreement for** *Alice* **about** *file* **with** print

and $agr'$ is

**agreement for** *Bob* **about** *file* **with** true $\longmapsto$ print.

Observe that $agr$ gives Alice permission to print the file and $agr'$ forbids Alice from printing it, since the agreement gives Bob the right exclusively. Because the agreements contradict each other, $f_q^+$ and $f_q^-$ are E-valid for all queries $q = (A, s, act, a, E)$. So the answer to the query $(A, Charlie, \mathsf{print}, file, E)$ is "Query inconsistent", whereas the answer to the query $(\{agr\}, Charlie, \mathsf{print}, file, E)$ and to the query $(\{agr'\}, Charlie, \mathsf{print}, file, E)$ is "Permission unregulated". ∎

If we consider only those queries in $\mathcal{Q}_1$ for which the set of agreements holds in at least one relevant model, then we get the desired results.

**Lemma 4.4.** *Suppose that $q = (A, s, act, a, E)$ is a query in $\mathcal{Q}_1$ such that $\bigwedge_{agr \in A} [\![agr]\!]$ is satisfied in at least one E-relevant model. For every $agr \in A$, let $q_{agr}$ be the query $(\{agr\}, s, act, a, E)$. Then:*

*(a) $f_q^+$ is E-valid if and only if $f_{q_{agr}}^+$ is E-valid for some $agr \in A$ and*

*(b) $f_q^-$ is E-valid if and only if $f_{q_{agr}}^-$ is E-valid for some $agr \in A$.*

It follows from Lemma 4.2 and 4.4 together that answering a query $q = (A, s, act, a, E)$ $\mathcal{Q}_1$ can be done efficiently, provided that $\bigwedge_{agr \in A} [\![agr]\!]$ is satisfied in at least one E-relevant model. Moreover, if this is not the case, then the query can be answered immediately. If $\bigwedge_{agr \in A} [\![agr]\!]$ does not hold in any E-relevant model then both $f_q^+$ and $f_q^-$ are E-valid, so the answer to $q$ is "Query inconsistent". Therefore, we can answer queries in $\mathcal{Q}_1$ efficiently provided we can quickly determine whether the agreements are satisfied in at least one relevant model.

**Lemma 4.5.** *There is an algorithm that, given a query $q = (A, s, act, a, E)$ in $\mathcal{Q}_1$, determines whether $\bigwedge_{agr \in A} [\![agr]\!]$ is satisfied in at least one E-relevant model in time $O(|E| |A|^8)$.*

*Proof.* See Appendix A. ∎

Putting all of these results together, we can derive the tractability of answering queries in $\mathcal{Q}_1$.

**Theorem 4.6.** *There is an algorithm that, given a query $q = (A, s, act, a, E)$ in $\mathcal{Q}_1$, computes the answer to $q$ in time $O(|E| |A|^8)$.*

*Proof.* First, run the algorithm of Lemma 4.5 to determine if $\bigwedge_{agr \in A} [\![agr]\!]$ is satisfied in at least one $E$-relevant model. This can be done in time $O(|E| \, |A|^8)$. If the result is "No", then return "Query inconsistent". If the result is "Yes", then use the algorithms of Lemma 4.2 to check whether $f_{q_{agr}}^+$ and $f_{q_{agr}}^-$ are $E$-valid for each query $q = (\{agr\}, s, act, a, E)$ such that $agr \in A$. This can be done in time $O(|E| \, |A|^7)$: there are less than $|A|$ agreements in $A$, and for every $agr \in A$, $|agr| \le |A|$. By Lemma 4.4, $f_q^+$ is $E$-valid if and only if $f_{q_{arg}}^-$ is $E$-valid for an $agr \in A$, and similarly for $f_q^-$. Thus, if $f_{q_{arg}}^+$ is $E$-valid for an $agr \in A$, and $f_{q_{arg}}^-$ is not $E$-valid for all $agr \in A$, then return "Permission granted". Similarly, if $f_{q_{arg}}^-$ is $E$-valid for an $agr \in A$, and $f_{q_{arg}}^+$ is not $E$-valid for all $agr \in A$, then return "Permission denied". Otherwise, return "Permission unregulated". ∎

We conclude this section with a few observations. First, we suspect that the queries that are of practical interest have certain properties that could be used to improve the efficiency of our algorithms. For example, it seems unlikely that a set of agreements will give one principal an exclusive right and give someone else that same right (possibly under certain conditions). That is, if $A$ is a set of agreements such that an agreement in $A$ gives a principal *prin* the exclusive-right to do an action *act* to an asset $a$ and another agreement in $A$ gives a principal *prin'* the right to do *act* to $a$ if certain prerequisites hold, then we expect that $subjects(prin') \subseteq subjects(prin)$. A straightforward syntactic check can be used to verify that this is indeed the case for a particular query and, if it is, then our proof of Theorem 4.5 can be easily modified to show that we can do the check in time $O(|E|)$.

We conjecture that answering a query $(A, s, act, a, E)$ in ODRL can be done efficiently, provided that, if an agreement in $A$ mentions a prerequisite of the form $\mathsf{xor}[prq_1, \ldots, prq_n]$, then $prq_i$ does not mention a prerequisite of the form $\mathsf{not}[ps]$, where $ps$ is a policy set, for $i = 1, \ldots, n$. That is, we suspect that answering queries can be done efficiently provided that, whether a permission holds, does not depend on whether a policy set holds (although it can depend on whether a policy set does not hold). We believe that we can use ideas discussed in [Halpern and Weissman 2003] to prove this result, however, we have not checked the details because, as previously discussed, it is not clear that such a result is of practical interest.

## 5 Discussion: Improving ODRL

The process of working through the ODRL specification to derive the formal semantics highlighted a number of potential weaknesses in the design of ODRL. In addition to not having formal semantics, the ODRL specification does not discuss which agreements should be enforced, how conflicts should be resolved, how agreements can be revoked, and how the environment can be maintained. We examine these issues in turn.

The ODRL specification does not say which agreements should be used when evaluating requests. The developers seem to assume that only a legitimate agent will be able to create a particular agreement; however, it is not clear which agents should be recognized as legitimate. Are there ODRL agreements that give subjects the right to create agreements? If so, who is allowed to write those agreements? A natural approach is simply to assume that everyone can write agreements; it is up to the enforcing system to determine which are legitimate. A problem with this design is that an agreement might be meaningless on some systems and quite significant on others. For example, suppose that Bob stores his diary on his home machine, which assumes all agreements are legitimate, and on his work machine, which assumes an agreement is only legitimate if written by a manager of the company. If Bob's sister Alice, who is not a manager of the company, writes an agreement that gives her permission to see Bob's diary, then the home machine will permit the access while the work machine will not.

A more satisfying approach is to define the circumstances under which an agreement is legitimate and require only legitimate agreements to be considered during query evaluation. A definition for legitimacy might say that some agreements are legitimate by fiat (e.g., any agreement about an asset $a$ issued by its owner), while others are legitimate because there is some proof of legitimacy (e.g., an agreement about an asset $a$ issued by subject $s$ is legitimate, because the owner of $a$ has written an agreement that gives $s$ permission to regulate access to $a$). This is essentially the approach adopted for XrML [ContentGuard 2001].

The ODRL specification does not discuss how conflicts should be resolved. For example, suppose that Alice gives Bob the exclusive right to distribute her movie and she gives Charlie the right to distribute it as well. Is Charlie allowed to distribute the movie? By the definition given in Section 4, the answer is "Query inconsistent" because the agreements are inconsistent in the environment (regardless of what the environment is). While this is an accurate description of the situation, it is not particularly helpful. One solution is to remove exclusive policy sets from the language, so that conflicts cannot occur. Another option is to store agreements with the relevant asset, rather than only with the users; that way, conflicts can be detected, and hopefully resolved between the relevant parties, as soon as a conflicting agreement is written. Finally, it is worth noting that, in languages such as XACML [Moses 2005] and FAF [Jajodia et al. 2001], conflicts are handled by requiring users to write overriding policies, such as "If an action is both permitted and forbidden, then it is forbidden". Unfortunately, it is not exactly clear how this solution could be incorporated into the ODRL framework.

The ODRL specification discusses revocation, but does not give a mechanism for revoking agreements or for checking whether an agreement has been revoked. Since prerequisites in ODRL can limit the time period in which a policy applies and the number of times the policy can be used to justify an action, it is not clear that revocation is truly necessary. Therefore, one solution is simply to remove all mention of revocation from the ODRL specification. Another option is to create policies under which an agreement can be revoked legitimately. These policies could be part of an agreement, or could be built-in to ODRL. The environment could then maintain a list of revoked agreements, which would not be used when answering queries.

Finally, the specification does not discuss how the environment is maintained. Holzer, Katzenbeisser, and Schallhart [2004] propose a solution to this problem. They associate with every ODRL agreement an automaton that transitions whenever the user of an agreement performs an action. Thus, to recast their work using our terminology, the states of the automaton corresponding to an agreement are what we call environments. Holzer et al. do not describe how to compute which actions are allowed in any given environment, however, they describe how to update the environment. In contrast, we do not describe how to update environments, but our semantics describes how to compute which actions are permitted in any given environment. In this sense, the two semantics are complementary.

# 6   Conclusion

ODRL is a popular rights language with features that we have not found in other approaches. However, the usefulness of ODRL is limited, in part, because the language does not have formal semantics. To address this deficiency, we have proposed a formal semantics for ODRL. In the process of creating this semantics, we discovered aspects of the specification that should be clarified and have discussed our findings with the language developers. They are currently working on the next version of the language, which has formal semantics as one of its seven design requirements.

In addition to giving the language formal semantics, we have considered the practical problem of determining whether a set of ODRL statements imply a permission or prohibition. Using our semantics, we have formally defined the problem and shown that it is, in general, NP-hard. By removing a component of ODRL whose meaning seems to be somewhat unclear, even to the developers, we can create a tractable fragment of the language. To prove that the fragment is tractable, we naturally created a polynomial-time algorithm to determine whether a set of ODRL statements imply a permission (or prohibition). To the best of our knowledge, this is the first algorithm for answering such queries in ODRL.

Despite these successes, the work is far from done. We are currently collaborating with the language developers on the next version of ODRL. We are also interested in examining other types of queries, such as what, if anything, can a subject do to get a desired permission. Finally, we intend to do a careful comparison of ODRL and a number of other languages in the near future.

### Acknowledgments

Thanks to Renato Iannella, who took the time to answer our questions about the interpretation of various ODRL components. We would also like to thank Joseph Halpern for his helpful comments on preliminary drafts. This work was supported in part by NSF under grant CTC-0208535, by ONR under grant N00014-01-10-511, by the DoD Multidisciplinary University Research Initiative (MURI) program administered by the ONR under grants N00014-01-1-0795 and N00014-04-1-0725, and by AFOSR under grant F49620-02-1-0101 and FA9550-05-1-0055.

# A   Proofs

In the proofs below, we use the notation $\mathcal{C}(S)$ for the cardinality of set $S$. We also use the notation $f[t/x]$ for the capture-avoiding substitution of term $t$ for variable $x$ in formula $f$.

**Theorem 4.1.** *The problem of deciding, for a query $q = (A, s, act, a, E)$, whether $f_q^+$ is E-valid is decidable and NP-hard. Similarly, the problem of deciding, for a query $q = (A, s, act, a, E)$, whether $f_q^-$ is E-valid is decidable.*

*Proof.* To prove decidability, we present an algorithm to determine whether $f_q^+$ is $E$-valid. The first step of the algorithm is to check if $E$ is inconsistent, by simply scanning $E$. (Recall that $E$ is inconsistent if and only if $E$ has two conjuncts of the form $count(s, id) = n$ and $count(s, id) = n'$ with $n \neq n'$.) If $E$ is inconsistent, then there are no $E$-relevant models, $f_q^+$ is trivially $E$-valid, and the algorithm returns "Yes".

If $E$ is consistent, then the set of $E$-relevant models is not empty, and the algorithm proceeds as follows. Let $g$ be the formula obtained from $f_q^+$ by replacing every subformula of the form $\forall x(h)$ by $\bigwedge_{s \in S}(h[s/x])$ and every subformula of the form $\exists x(h)$ by $\bigvee_{s \in S}(h[s/x])$, where $S$ is the set of variable-free terms mentioned in $q$ that are the same sort as $x$. We claim that $f_q^+$ is $E$-valid if and only if $g$ is $E$-valid. We prove this claim by progressively constructing $g$; during this process, we consider in some detail the subformulas of the form $\forall x(h)$ and $\exists x(h)$ that can appear in $f_q^+$.

- Let $g_0$ be the formula obtained from $f_q^+$ by replacing every subformula of the form

$$\forall x((x = s_1 \vee \ldots \vee x = s_n \wedge g') \Rightarrow (g'' \Rightarrow \mathbf{Permitted}(x, act', a')))$$

  by

$$\bigwedge_{s \in S}((x = s_1 \vee \ldots \vee x = s_n \wedge g') \Rightarrow (g'' \Rightarrow \mathbf{Permitted}(x, act', a')))[s/x],$$

  where $S$ is the set of variable-free terms of sort *Subjects* mentioned in $q$. Since $\{s_1, \ldots, s_n\} \subseteq S$, it is easy to see that $f_q^+$ is $E$-valid if and only if $g_0$ is $E$-valid.

- Let $\Sigma$ be the set of substitutions $\sigma$ such that, for all variables $t$ of sort *Times* in $g_0$, $\sigma(t)$ is a variable-free term of sort *Times* that appears in $q$ and, for all other variables $x$, $\sigma(x) = x$. Note that $\Sigma$ is finite. Let $g_1$ be the formula obtained from $g_0$ by replacing every formula of the form $\exists t_1 \ldots \exists t_n(h)$, where every free variable of $h$ is of sort *Times*, with $\bigvee_{\sigma \in \Sigma}(h\sigma)$. It follows from the translation that, if $t$ is a free variable in $h$, then $h$ is a conjunction of formulas and one of those conjuncts has either the form $\mathbf{Paid}(r, I, t)$ or the form $\mathbf{Attributed}(s, t)$. It follows from the closed-world assumption that $g_0$ is $E$-valid if and only if $g_1$ is $E$-valid.

- It follows from the translation that every variable remaining in $g_1$ is of sort *Subjects*; $g_1$ includes a subformula of the form $\forall x(h)$ if and only if $h$ can be written as $x \neq s_1 \wedge \ldots \wedge x \neq s_n \Rightarrow \neg\mathbf{Permitted}(x, act', a')$, where $s_i$ is a variable-free term in $q$, for $i = 1, \ldots, n$. Let $g_2$ be the formula obtained from $g_1$ by replacing every subformula of the form $\forall x(h)$ by $\bigwedge_{s \in S} h[s/x]$, where $S$ is the set of variable-free terms of sort *Subjects* mentioned in $q$. Note that $g_2 = g$. So, it remains to show that $g_1$ is $E$-valid if and only if $g_2$ is $E$-valid. The "if" direction is trivial. For the "only if" direction, suppose by way of contradiction that $g_1$ is $E$-valid and $g_2$ is not. Note that $g_1$ is of the form $g_1' \Rightarrow \mathbf{Permitted}(s, act, a)$ and $g_2$ is of the form $g_2' \Rightarrow \mathbf{Permitted}(s, act, a)$ for appropriate formulas $g_1'$ and $g_2'$. Since $g_2$ is not $E$-valid, there is an $E$-relevant model $M$ that satisfies $g_2' \wedge \neg\mathbf{Permitted}(s, act, a)$. Let $M'$ be the $E$-relevant model that is identical to $M$, except that the domain of $M'$ is limited to the closed terms that are mentioned in $q$. It is easy to see that $g_2'$ holds in $M'$ since the formula holds in $M$, is variable-free, and mentions only those terms that appear in $q$. It follows from the construction of $g_2$ that, because $g_2'$ holds in $M'$, $g_1'$ holds in $M'$. Since, by construction, $M'$ does not satisfy **Permitted**$(s, act, a)$, $M$ does not satisfy $g_1$, which gives us the desired contradiction.

Since $g$ is variable-free, the algorithm proceeds by replacing every **Permitted**-free literal appearing in $g$ by either **true** or **false** depending on $E$ and the standard interpretations of $=$, $<$ and $\leq$. Let $h$ be the formula obtained from $g$ by doing this replacement. Clearly, $g$ is $E$-valid if and only if $h$ is $E$-valid. Moreover, since **Permitted** is the only predicate symbol appearing in $h$, $h$ is $E$-valid if and only if $h$ is valid. The algorithm determines the validity of $h$ by checking if $h$ holds for all assignments of **true** or **false** to the **Permitted** literals in $h$ (where a positive literal $\ell$ is not given the same assignment as $\neg\ell$). Obviously, $h$ is valid if it holds under every substitution and is not valid otherwise.

The same strategy can be used to derive an algorithm that determines the $E$-validity of $f_q^-$.

We now reduce the 3-satisfiability problem to the problem of determining whether $f_q^+$ is $E$-valid for an appropriate query $q$, thereby showing that the latter problem is NP-hard. Let $\varphi = C_1 \wedge \ldots \wedge C_n$ be a formula in propositional logic, where each $C_i$ is a clause with three disjuncts. Without loss of generality, we assume that no conjunct $C_i$ is valid. Let $P_1, \ldots P_m$ be the primitive propositions mentioned in $\varphi$. We want to determine if $\varphi$ is satisfiable.

Let $s_0, \ldots, s_m$ be subjects and let $a$ be an asset. For each conjunct $C_i = L_1 \vee L_2 \vee L_3$ of $\varphi$, let $agr_i$ be the agreement

**agreement for** $\{s_0, \ldots, s_m\}$ **about** $a$ **with** $\mathsf{and}[prq_1, prq_2, prq_3] \Rightarrow \mathsf{display}$,

where

$$prq_j \triangleq \begin{cases} \mathsf{and}[s_0, \mathsf{not}[s_k \Rightarrow \mathsf{print}]] & \text{if } L_j \text{ is } P_k \\ \mathsf{and}[s_0, \mathsf{xor}[\mathsf{true}, \mathsf{not}[s_k \Rightarrow \mathsf{print}]]] & \text{if } L_j \text{ is } \neg P_k. \end{cases}$$

Let $q$ be the query $(\{agr_1, \ldots, agr_n\}, s_0, \mathsf{display}, a, E)$, where $E$ is the empty environment (i.e., **true**). We claim that $\varphi$ is satisfiable if and only if $f_q^+$ is not $E$-valid. For every assignment $A$ of truth values to $P_1, \ldots, P_m$, let $M_A$ be the $E$-relevant model that satisfies $\neg$**Permitted**$(s_i, print, a)$ if and only if $A$ assigns $P_i$ to **false** or $s_i = 0$. It is not hard to show that a truth assignment $A$ satisfies a conjunct $c_i$ of $\varphi$ if and only if $M_A$ satisfies $[\![agr_i]\!]$. The key observation is that, for each conjunct $C_i = L_1 \vee L_2 \vee L_3$ of $\varphi$, we can write $[\![agr_i]\!]$ as

$$f_{i,1} \wedge f_{i,2} \wedge f_{i,3} \Rightarrow \textbf{Permitted}(s_0, display, a),$$

where

$$f_{i,j} = \begin{cases} \neg\textbf{Permitted}(s_k, print, a) & \text{if } L_j \text{ is } P_k \\ \textbf{Permitted}(s_k, print, a) & \text{if } L_j \text{ is } \neg P_k. \end{cases}$$

So, if $\varphi$ is satisfiable, then there is a truth assignment $A$ that satisfies $\varphi$, the model $M_A$ satisfies $\bigwedge_{agr \in A}[\![agr]\!] \wedge \neg\textbf{Permitted}(s_0, display, a)$, and, thus, $f_q^+$ is not $E$-valid. If $\varphi$ is not satisfiable then, for every truth assignment $A$, $M_A$ does not satisfy some $[\![agr_i]\!]$, so $M_A$ satisfies $f_q^+$. Let $\mathcal{M}$ be the set of models $M$ such that, for all truth assignments $A$, $M \neq M_A$. It is not hard to see that every model in $\mathcal{M}$ satisfies **Permitted**$(s_0, display, a)$, thereby satisfying $f_q^+$. Since every $E$-relevant model satisfies $f_q^+$, the formula is $E$-valid. ∎

The following result is used in Lemmas 4.2 and 4.4.

**Lemma A.1.** *Suppose that $f$ is a **Permitted**-free formula and $E$ is an environment such that the set of $E$-relevant models is nonempty. Then $f$ holds in at least one $E$-relevant model if and only if $f$ is $E$-valid.*

*Proof.* Follows immediately from the definitions. ∎

Given a policy set $ps$, let $S_{ps}^+$ be the set of tuples $(prq, I, prq', id, act')$ such that $ps$ mentions the policy set $prq \longrightarrow p$ or $prq \longmapsto p$, $I$ is the set of policy identifiers appearing in $p$, and $p$ mentions the policy $prq' \Longrightarrow_{id} act'$. Finally, let $S_{ps}^-$ be the set of actions such that an action $act'$ is in $S_{ps}^-$ if and only if $ps$ mentions an exclusive policy set that mentions a policy of the form $prq \Longrightarrow act'$.

**Lemma A.2.** *Suppose $agr$ is an agreement of the form*

$$\textbf{agreement for } prin_u \textbf{ about } a \textbf{ with } ps.$$

*Then $[\![agr]\!]$ holds in model $M$ if and only if*

(a) *for every $act' \in S_{ps}^-$ and $s' \notin subjects(prin_u)$, $M$ satisfies $\neg\textbf{Permitted}(s', act', a)$, and*

(b) *for every $(prq, I, prq', id, act') \in S_{ps}^+$ and $s' \in subjects(prin_u)$, either $M$ satisfies $\textbf{Permitted}(s', act', a)$ or $M$ does not satisfy $[\![prq]\!]_{s'}^{I, prin_u} \wedge [\![prq']\!]_{s'}^{\{id\}, prin_u}$.*

*Proof.* Immediate by the definition of $S_{ps}^+$ and $S_{ps}^-$ and the translation $[\![\cdot]\!]$. ∎

**Lemma 4.2.** *There are algorithms that, given a query $q = (\{agr\}, s, act, a, E)$ in $\mathcal{Q}_1$:*

(a) *determine whether $f_q^+$ is $E$-valid in time $O(|E|\,|agr|^6)$, and*

(b) *determine whether $f_q^-$ is $E$-valid in time $O(|E| + |agr|)$.*

*Proof.* Suppose that $agr$ is an agreement of the form **agreement for** $prin_u$ **about** $a'$ **with** $ps$.

For part (a), we claim that $[\![agr]\!] \Rightarrow \textbf{Permitted}(s, act, a)$ is $E$-valid if and only if the set of $E$-relevant models is empty, or all of the following conditions hold:

(i) $s \in subjects(prin_u)$,

(ii) $a' = a$, and

(iii) there is a tuple $(prq, I, prq', id, act) \in S_{ps}^+$ such that $[\![prq]\!]_s^{I, prin_u} \wedge [\![prq']\!]_s^{\{id\}, prin_u}$ is $E$-valid.

- For the "if" direction, if the set of $E$-relevant models is empty, then the formula $[\![agr]\!] \Rightarrow \textbf{Permitted}(s, act, a)$ is trivially $E$-valid. If (i), (ii), and (iii) hold, then it is immediate from the translation that $[\![agr]\!] \Rightarrow \textbf{Permitted}(s, act, a)$ is $E$-valid.

- For the "only if" direction, suppose by way of contradiction that the formula $[\![agr]\!] \Rightarrow \textbf{Permitted}(s, act, a)$ is $E$-valid, the set of $E$-relevant models is not empty, and either (i), (ii), or (iii) does not hold. Because the set of $E$-relevant models is not empty, there is a model $M$ that is $E$-relevant and that satisfies $\textbf{Permitted}(t_1, t_2, t_3)$ if and only if $t_1 \in subjects(prin_u)$, $t_3 = a'$, and $\textbf{Permitted}(t_1, t_2, t_3) \neq \textbf{Permitted}(s, act, a)$, for all closed terms $t_1$, $t_2$, and $t_3$ of the appropriate sorts. We claim that $M$ satisfies $[\![agr]\!]$, thus contradicting the assumption that $[\![agr]\!] \Rightarrow \textbf{Permitted}(s, act, a)$ is $E$-valid. By Lemma A.2, it suffices to show that A.2(a) and A.2(b) hold. A.2(a) follows from the construction of $M$. If (i) or (ii) does not hold, then $M$ satisfies $\textbf{Permitted}(s', act', a')$, for every tuple $(prq, I, prq', I', act')$ in $S_{ps}^+$, so A.2(b) holds. Suppose that (iii) does not hold. Then, for each tuple $(prq, I, prq', I', act') \in S_{ps}^+$ and subject $s' \in subjects(prin_u)$, either $s' \neq s$, in which case $M$ satisfies $\textbf{Permitted}(s', act', a')$; $act' \neq act$, in which case $M$ satisfies $\textbf{Permitted}(s', act', a')$; or $s' = s$, $act' = act$, and $f = [\![prq]\!]_s^{I, prin_u} \wedge [\![prq']\!]_s^{\{id\}, prin_u}$ is not $E$-valid. It follows from Lemma A.1 that $f$ does not hold in $M$ because it is $\textbf{Permitted}$-free (neither $prq$ nor $prq'$ mention a policy set), so A.2(b) holds again.

It follows that we can determine the $E$-validity of $[\![agr]\!] \Rightarrow \textbf{Permitted}(s, act, a)$ by running the following algorithm: determine whether the set of $E$-relevant models is empty; if so, return "Yes", otherwise check conditions (i), (ii), and (iii); if all hold, then return "Yes", else return "No". The set of $E$-relevant models is non-empty if and only if $E$ is inconsistent, which can be checked in time $O(|E|)$. We can check whether (i) and (ii) hold in time $O(|agr|)$. We can also compute $S_{ps}^+$ in time $O(|agr|)$. Finally, the cardinality of $S_{ps}^+$ is less than $|agr|$. We show that, for each tuple $(prq, I, prq', id, act)$ in $S_{ps}^+$, we can determine whether $[\![prq]\!]_s^{I, prin_u} \wedge [\![prq']\!]_s^{\{id\}, prin_u}$ is $E$-valid in time $O(|E| \, |agr|^5)$, so the total runtime of the algorithm is $O(|E| \, |agr|^6)$.

Using the translation as a guide, we can construct an algorithm for determining whether $[\![prq]\!]_s^{I, prin_u}$ (or $[\![prq']\!]_s^{\{id\}, prin_u}$) is $E$-valid. The first step is to rewrite the prerequisites $prq$ and $prq'$ so that they do not contain nested forEachMember constraints. Examining the translation, it is clear that the constraint

$$\mathsf{forEachMember}[prin; \mathsf{forEachMember}[prin'; cons'], cons]$$

translates to a formula that is logically equivalent to the translation of

$$\mathsf{and}[\mathsf{forEachMember}[prin; cons], \mathsf{forEachMember}[prin'; cons']].$$

Generalizing this idea, we can rewrite, in time $O(|prq|)$, the prerequisite $prq$ to an equivalent $prq_0$ of size $O(|prq|)$ that does not contain nested forEachMember constraints, and similarly rewrite $prq'$ to an equivalent $prq'_0$. We then apply the algorithm given in Figure 5, called $Holds$, to $prq_0$ and $prq'_0$. The algorithm $Holds$ returns $\textbf{true}$ or $\textbf{false}$; it calls $ReqHolds$, which is given in Figure 6, and which returns the earliest time at which a given requirement holds, or $\textbf{false}$ if the requirement never holds. The claim that $Holds(prq, s, I, prin_u, E) = \textbf{true}$ if and only if $[\![prq]\!]_s^{I, prin_u}$ is $E$-valid is established by a straightforward induction on the structure of $prq$. We can check that the algorithm runs in time $O(|E| \, |agr|^5)$ by solving a simple recurrence equation. (The assumption that there are no nested forEachMember

in $prq$ is crucial to obtain this running time; without this assumption, the running time is exponential in the size of the prerequisite.) We leave the straightforward details to the reader.

For part (b), we claim that $[\![agr]\!] \Rightarrow \neg\mathbf{Permitted}(s, act, a)$ is $E$-valid if and only if the set of $E$-relevant models is empty or all of the following conditions hold:

(i) $s \notin subjects(prin_u)$,

(ii) $a' = a$, and

(iii) $agr$ includes an exclusive policy set that mentions a policy of the form $prq \Rightarrow act$.

- For the "if" direction, if the set of $E$-relevant models is empty, then the formula $[\![agr]\!] \Rightarrow \neg\mathbf{Permitted}(s, act, a)$ is trivially $E$-valid. If (i), (ii), and (iii) hold then $[\![agr]\!]$ can be written as a conjunction of formulas, one of which says that every subject who is not mentioned in $prin_u$ is forbidden to do $act$ to $a$, so $[\![agr]\!] \Rightarrow \neg\mathbf{Permitted}(s, act, a)$ is again $E$-valid.

- For the "only if" direction, suppose that the set of $E$-relevant models is non-empty. It follows that there is an $E$-relevant model $M$ such that, for all closed terms $t_1$, $t_2$, and $t_3$ of the appropriate sorts, $M$ satisfies $\neg\mathbf{Permitted}(t_1, t_2, t_3)$ if and only if $t_1 \notin subjects(prin_u)$, $t_3 = a'$, and $\neg\mathbf{Permitted}(t_1, t_2, t_3) \neq \neg\mathbf{Permitted}(s, act, a)$. We claim that, if (i), (ii), or (iii) does not hold, then $[\![agr]\!]$ holds in $M$ and, thus, $[\![agr]\!] \Rightarrow \neg\mathbf{Permitted}(s, act, a)$ is not $E$-valid. By Lemma A.2, it suffices to show that A.2(a) and A.2(b) hold. Since $M$ satisfies $\mathbf{Permitted}(t_1, t_2, t_3)$ for all closed terms such that $t_1 \in subjects(prin_u)$, A.2(b) holds. If (i) or (ii) does not hold, then $M$ satisfies $\neg\mathbf{Permitted}(t_1, t_2, t_3)$ if and only if $t_1 \notin subjects(prin_u)$ and $t_3 = a'$. It follows that, for all subjects $s' \notin subjects(prin_u)$ and all actions $act'' \in S_{ps}^-$, $M$ satisfies $\neg\mathbf{Permitted}(s', act'', a')$; so A.2(a) holds. If (iii) does not hold, then $act \notin S_{ps}^-$. It follows from the construction of $M$ that, for each action $act'' \neq act$ and each subject $s' \notin subjects(prin_u)$, $M$ satisfies $\neg\mathbf{Permitted}(s', act'', a')$, so A.2(b) holds.

Thus, we can determine the $E$-validity of $[\![agr]\!] \Rightarrow \neg\mathbf{Permitted}(s, act, a)$ by running the following algorithm: determine whether the set of $E$-relevant models is empty; if so, return "Yes", otherwise check conditions (i), (ii), and (iii); if all hold, then return "Yes", else return "No". Checking that the set of $E$-relevant models is empty can be done in time $O(|E|)$. Checking conditions (i), (ii), and (iii) can be done in time $O(|A|)$. ∎

**Lemma 4.4.** *Suppose that* $q = (A, s, act, a, E)$ *is a query in* $\mathcal{Q}_1$ *such that* $\bigwedge_{agr \in A}[\![agr]\!]$ *is satisfied in at least one $E$-relevant model. For every* $agr \in A$*, let* $q_{agr}$ *be the query* $(\{agr\}, s, act, a, E)$*. Then:*

(a) $f_q^+$ *is $E$-valid if and only if* $f_{q_{agr}}^+$ *is $E$-valid for some* $agr \in A$*, and*

(b) $f_q^-$ *is $E$-valid if and only if* $f_{q_{agr}}^-$ *is $E$-valid for some* $agr \in A$*.*

$Holds(prq, s, I, prin_u, E) \triangleq$

   if $prq = \mathsf{true}$ then return **true**

   if $prq = prin$ then
      if $s \notin subjects(prin)$ then return **true** else return **false**

   if $prq = \mathsf{forEachMember}[prin, cons_1, \ldots, cons_m]$ then
      if $Holds(cons_i, S, I, prin', E)$ is **true**
         for all $i = 1, \ldots, m$ and all $prin' \in subjects(prin)$ then
      return **true**
     else return **false**

   if $prq = \mathsf{count}[n]$ then
     $sum := 0$
     for each $s' \in subjects(prin_u)$
       for each $id \in I$
         if $count(s, id) = n'$ is a conjunct of $E$ then
           $sum := sum + n'$
     if $sum < n$ then return **true** else return **false**

   if $prq = prin\langle\mathsf{count}[n]\rangle$ then return $Holds(\mathsf{count}[n], s, I, prin, E)$

   if $prq = \mathsf{not}[cons]$ then return $\neg Holds(cons, s, I, prin_u, E)$

   if $prq = \mathsf{and}[prq_1, \ldots, prq_m]$ then return $\wedge_{i=1}^{m} Holds(prq_i, s, I, prin_u, E)$

   if $prq = \mathsf{or}[prq_1, \ldots, prq_m]$ then return $\vee_{i=1}^{m} Holds(prq_i, s, I, prin_u, E)$

   if $prq = \mathsf{xor}[prq_1, \ldots, prq_m]$ then
     $seenone := $ **false**
     for $i = 1, \ldots, m$
       if $Holds(prq_i, s, I, prin_u, E)$ is **true** and $seenone$ is **false** then
         $seenone := $ **true**
       if $Holds(prq_i, s, I, prin_u, E)$ is **true** and $seenone$ is **true** then
         return **false**
     return $seenone$

   if $prq$ is a requirement then return $ReqHolds(prq, s, I, prin_u, E, 0, \infty) \neq$ **false**

Figure 5: Algorithm $Holds$

$RegHolds(req, s, I, prin_u, E, t, t_{max}) \triangleq$

    if $req = \mathsf{prePay}[r]$ then
       $t' := t_{max}$
       for each conjunct $\ell$ of $E$
           if $\ell$ is of the form $\mathbf{Paid}(r, I, t'')$ and $t \leq t'' < t'$ then
              $t' := t''$
       if $t' \neq t_{max}$ then return $t'$ else return $\mathbf{false}$

    if $prq = \mathsf{attribution}[s]$ then
       $t' := t_{max}$
       for each conjunct $\ell$ of $E$
           if $\ell$ is of the form $\mathbf{Attributed}(s, t'')$ and $t \leq t'' < t'$ then
              $t' := t''$
       if $t' \neq t_{max}$ then return $t'$ else return $\mathbf{false}$

    if $prq = \mathsf{anySeq}[req_1, \ldots, req_m]$ then
       $t' := RegHolds(req_1, s, I, prin_u, E, t, t_{max})$
       if $t' \neq \mathbf{false}$ then return $RegHolds(\mathsf{anySeq}[req_2, \ldots, req_m], s, I, prin_u, E, t, t_{max})$
       else return $\mathbf{false}$

    if $prq = \mathsf{inSeq}[req_1, \ldots, req_m]$ then
       $t' := RegHolds(req_1, s, I, prin_u, E, t, t_{max})$
       if $t'$ is $\mathbf{false}$ then return $\mathbf{false}$
       else return $RegHolds(\mathsf{inSeq}[req2, \ldots, req_m], s, I, prin_u, E, t', t_{max})$

Figure 6: Algorithm *RegHolds*

*Proof.* For part (a), the "if" direction is trivial. For the "only if" direction, suppose by way of contradiction that $\bigwedge_{agr \in A} [\![agr]\!] \Rightarrow \mathbf{Permitted}(s, act, a)$ is $E$-valid and $[\![agr]\!] \Rightarrow \mathbf{Permitted}(s, act, a)$ is not $E$-valid for every $agr \in A$. By assumption, there is an $E$-relevant model $M$ that satisfies $\bigwedge_{agr \in A} [\![agr]\!]$. Let $M'$ be the model that is identical to $M$ except that $M'$ satisfies $\neg\mathbf{Permitted}(s, act, a)$. Because $M$ is $E$-relevant and $M'$ differs from $M$ only on the interpretation of $\mathbf{Permitted}$, $M'$ is $E$-relevant. Since $M'$ satisfies $\neg\mathbf{Permitted}(s, act, a)$ and, by assumption, $\bigwedge_{agr \in A} [\![agr]\!] \Rightarrow \mathbf{Permitted}(s, act, a)$ is $E$-valid, there is an agreement $agr$ in $A$ such that $M'$ does not satisfy $[\![agr]\!]$. We now show that $[\![agr]\!]$ implies $\mathbf{Permitted}(s, act, a)$, which contradicts the assumptions. Because no agreement in $A$ mentions a condition of the form $\mathsf{not}[ps]$, it follows from the translation that we can write $[\![agr]\!]$ as $\forall x(f_1) \wedge \cdots \wedge \forall x(f_n)$, where each $f_i$ is of the form $g \Rightarrow (\neg)\mathbf{Permitted}(x, act', a')$, $g$ is $\mathbf{Permitted}$-free, and both $act'$ and $a'$ are closed terms of the appropriate sorts. Because $[\![agr]\!]$ holds in $M$ and does not hold in $M'$, there exists integer $i$ such that $f_i = g \Rightarrow \mathbf{Permitted}(x, act, a)$ and $g[s/x]$ is satisfied in $M'$. Since $g[s/x]$ is $\mathbf{Permitted}$-free and is satisfied in a $E$-relevant model, it follows from Lemma A.1 that $g[s/x]$ is $E$-valid. Putting the pieces together, we can write $[\![agr]\!]$ as $\forall x(h \wedge (g \Rightarrow \mathbf{Permitted}(x, act, a)))$, for an appropriate formula $h$, and $g[s/x]$ is $E$-valid. It readily follows that $[\![agr]\!] \Rightarrow \mathbf{Permitted}(s, act, a)$ is $E$-valid.

The proof for part (b) is nearly identical to the proof for part (a); in fact, the former can be obtained from the latter by replacing every occurence of $\mathbf{Permitted}$ by $\neg\mathbf{Permitted}$ and vice versa. ∎

**Lemma 4.5.** *There is an algorithm that, given a query $q = (A, s, act, a, E)$ in $\mathcal{Q}_1$, determines whether $\bigwedge_{agr \in A} [\![agr]\!]$ is satisfied in at least one $E$-relevant model in time $O(|E|\,|A|^8)$.*

*Proof.* We claim that $\bigwedge_{agr \in A} [\![agr]\!]$ holds in an $E$-relevant model if and only if

(i) the set of $E$-relevant models is not empty, and

(ii) for every pair of agreements

> **agreement for** $prin_u$ **about** $a$ **with** $ps$, and
> **agreement for** $prin'_u$ **about** $a'$ **with** $ps'$

in $A$, either

(a) $a \neq a'$, or

(b) for all actions $act \in S_{ps}^-$, tuples $(prq, I, prq', id, act) \in S_{ps'}^+$, and subjects $s \in subjects(prin'_u) \backslash subjects(prin_u)$, $[\![prq]\!]_s^{I, prin'_u} \wedge [\![prq']\!]_s^{\{id\}, prin'_u}$ is not $E$-valid.

For the "if" direction, observe that if (i) holds, then there is an $E$-relevant model $M$ such that, for all closed terms $t_1, t_2$, and $t_3$ of the appropriate sort, $M$ satisfies $\neg\mathbf{Permitted}(t_1, t_2, t_3)$ if and only if there is an agreement $agr$ of the form **agreement for** $prin_u$ **about** $a$ **with** $ps$ in $A$ such that $t_1 \notin subjects(prin_u)$, $ps$ includes an exclusive policy set that mentions a policy of the form $prq \Rightarrow t_2$, and $t_3 = a$. It is not hard to see that, if (ii) holds, then $M$ satisfies $\bigwedge_{agr \in A} [\![agr]\!]$ and we are done. For the "only if" direction, observe that, if (i) does not hold, then $\bigwedge_{agr \in A} [\![agr]\!]$ clearly does not hold in an $E$-relevant model. If

(ii) does not hold, then there is a subject $s$, action $act$, and asset $a$, such that, for an agreement $agr \in A$, $[\![agr]\!] \Rightarrow \mathbf{Permitted}(s, act, a)$ is $E$-valid and, for an agreement $agr' \in A$, $[\![agr]\!] \Rightarrow \neg\mathbf{Permitted}(s, act, a)$ is $E$-valid. Since no model can satisfy both $\mathbf{Permitted}(s, act, a)$ and $\neg\mathbf{Permitted}(s, act, a)$, no $E$-relevant model can satisfy both $[\![agr]\!]$ and $[\![agr']\!]$, so $\bigwedge_{agr \in A}[\![agr]\!]$ does not hold in any $E$-relevant model.

We can determine whether (i) holds in time $O(|E|)$, since (i) holds if and only if $E$ is consistent. To check whether (ii) holds, we first construct the sets $S_{ps}^+$ and $S_{ps}^-$, which takes time $O(|A|)$; then we compare all $|A|^2$ pairs of agreements. For every agreement $agr$ in every pair of agreements, we determine whether certain prerequisites hold; this takes time $O(|E|\,|agr|^6)$, because there are at most $|agr|$ prerequisites per agreement $agr$ and evaluating each requirement takes time $O(|E|\,|agr|^5)$, as shown in the proof of Theorem 4.2. Since $|agr| \leq |A|$ for every agreement $agr \in A$, we get a total running time of $O(|E|\,|A|^8)$. ∎